\documentclass[11pt]{article}
\usepackage{epsfig}
\usepackage{graphicx,amssymb,amsmath}
\usepackage{latexsym}
\usepackage[dvips]{color}
\date{August 3, 2013 \\[0.5cm]
{ White Paper for the 2013 US HEP Community Summer Study (Snowmass2013)}}
\author{V.I.~Telnov\\[3mm]
{  Budker Institute of Nuclear Physics, 630090, Novosibirsk, Russia} \\
{  Novosibirsk State University, 630090, Novosibirsk, Russia }}
\title{\bf Comments on photon colliders for Snowmass 2013}

\begin{document}

\maketitle

\newcommand{\M}{\mbox{m}}
\newcommand{\n}{\mbox{$n_f$}}
\newcommand{\G}{\mbox{ee}}
\newcommand{\EP}{\mbox{$e^+$}}
\newcommand{\EM}{\mbox{$e^-$}}
\newcommand{\EPEM}{\mbox{$e^+e^{-}$}}
\newcommand{\EMEM}{\mbox{$e^-e^-$}}
\newcommand{\GG}{\mbox{$\gamma\gamma$}}
\newcommand{\GE}{\mbox{$\gamma e$}}
\newcommand{\GP}{\mbox{$\gamma e^+$}}
\newcommand{\TEV}{\mbox{TeV}}
\newcommand{\GEV}{\mbox{GeV}}
\newcommand{\LGG}{\mbox{$L_{\gamma\gamma}$}}
\newcommand{\LGE}{\mbox{$L_{\gamma e}$}}
\newcommand{\LEE}{\mbox{$L_{ee}$}}
\newcommand{\LEPEM}{\mbox{$L_{e^+e^-}$}}
\newcommand{\WGG}{\mbox{$W_{\gamma\gamma}$}}
\newcommand{\WGE}{\mbox{$W_{\gamma e}$}}
\newcommand{\EV}{\mbox{eV}}
\newcommand{\CM}{\mbox{cm}}
\newcommand{\MM}{\mbox{mm}}
\newcommand{\NM}{\mbox{nm}}
\newcommand{\MKM}{\mbox{$\mu$m}}
\newcommand{\SEC}{\mbox{s}}
\newcommand{\CMS}{\mbox{cm$^{-2}$s$^{-1}$}}
\newcommand{\MRAD}{\mbox{mrad}}
\newcommand{\IND}{\hspace*{\parindent}}
\newcommand{\E}{\mbox{$\epsilon$}}
\newcommand{\EN}{\mbox{$\epsilon_n$}}
\newcommand{\EI}{\mbox{$\epsilon_i$}}
\newcommand{\ENI}{\mbox{$\epsilon_{ni}$}}
\newcommand{\ENX}{\mbox{$\epsilon_{nx}$}}
\newcommand{\ENY}{\mbox{$\epsilon_{ny}$}}
\newcommand{\EX}{\mbox{$\epsilon_x$}}
\newcommand{\EY}{\mbox{$\epsilon_y$}}
\newcommand{\BI}{\mbox{$\beta_i$}}
\newcommand{\BX}{\mbox{$\beta_x$}}
\newcommand{\BY}{\mbox{$\beta_y$}}
\newcommand{\SX}{\mbox{$\sigma_x$}}
\newcommand{\SY}{\mbox{$\sigma_y$}}
\newcommand{\SZ}{\mbox{$\sigma_z$}}
\newcommand{\SI}{\mbox{$\sigma_i$}}
\newcommand{\SIP}{\mbox{$\sigma_i^{\prime}$}}
\newcommand{\be}{\begin{equation}}
\newcommand{\ee}{\end{equation}}
\newcommand{\bc}{\begin{center}}
\newcommand{\ec}{\end{center}}
\newcommand{\bi}{\begin{itemize}}
\newcommand{\ei}{\end{itemize}}
\newcommand{\ben}{\begin{enumerate}}
\newcommand{\een}{\end{enumerate}}
\newcommand{\bm}{\boldmath}

\begin{abstract}
  For more than 30 years~\cite{GKST81}, \GG, \GE\ photon colliders have been considered
a natural addition to \EPEM\ linear-collider projects.
Following the recent discovery of the Higgs boson, the physics community has been
actively considering various approaches to building a Higgs factory,
a photon collider (with or without \EPEM) being one of them.
In this note, following a brief discuss of photon colliders based on ILC and CLIC,
I give a critical overview of the recently proposed photon-collider
Higgs factories with no \EPEM\ collision option.

\end{abstract}

\section{Physics motivation}

In short, the photon collider can study New Physics at
energies and statistics similar to those in \EPEM collisions---but in
different reactions. In some cases, the photon collider provides access to higher
masses or allows the study of some phenomena with higher precision than.

Let us compare the strengths of \EPEM\ and \GG\ colliders in the study of the Higgs boson.
The photon collider can measure $\Gamma(H\to \GG)\times Br(H\to bb,ZZ,WW,\GG)$
and, using linearly polarized photons, the Higgs' $CP$ properties.
In order to extract $\Gamma(H \to \GG$), one needs the value of $Br(H \to
bb)$  from an \EPEM\ collider. In \EPEM\ collisions, one can measure $Br(H\to bb,
cc, gg, WW, ZZ, \mu\mu, \mathrm{invisible}), \Gamma_{\mathrm{tot}}$. The process $\EPEM \to ZH$ with $Z$ tagging
allows the measurement of the absolute values of branching fractions, including Higgs decays
to $\tau\tau, \mu\mu, cc$, which are not accessible in \GG\ collisions due to a large QED background.

  The rate of Higgs boson production in \GG\ collisons~\cite{TESLATDR}
  \be
  \dot{N}_H=L_{ee} \times \frac{dL_{0,\GG}}{dW_{\GG}L_{ee}}\frac{4\pi^2\Gamma_{\GG}}{M_H^2}(1+\lambda_1\lambda_2+CP*l_1l_2 cos2\varphi)=
  L_{ee}\sigma
  \ee
  $$    \sigma=\frac{0.98\cdot 10^{-35}}{2E_0[\GEV]} \frac{dL_{0,\GG}}{dz L_{ee}} (1+\lambda_1\lambda_2+CP*l_1l_2 cos2\varphi), \; \mbox{cm}$$
where $L_{ee}$ is the geometric $ee$ luminosity, $L_{0,\GG}$ is the \GG\ luminosity at total helicity zero,
$z=W_{\GG}/2E_0$, $\lambda_{1,2}$ and  $l_{1,2}$ are the helicities and linear polarizations of the high-energy photons,
$\varphi$ is the angle between the directions of linear polarizations, and $CP$ is the $CP$ parity of the Higgs boson.

The most reasonable choice of photon collider energy and the laser wavelength for the Higgs study is $E_0=110$ GeV
and $\lambda \sim 1.05$ \MKM\ (most powerful lasers available); the corresponding parameter $x=4E_0 \omega_0/m^2c^4 \approx 2$.

Let us consider the two most important sets of parameters: 1) for the measurement of $\Gamma_{\GG}$, 2) for the measurement of $CP$.
In both cases, it is preferable to use longitudinally polarized electrons, $2\lambda_e=-0.85$ is possible.
For case 1, the laser polarization should be $P_{\mathrm{c}}=1$ and $2P_{\mathrm{c}}\lambda_e \sim -0.85$
(to enhance the number of high-energy photons); then, the resulting  polarization of the scattered photons $\lambda_{1,2} \approx 1$, $l_{1,2}=0$.
For case 2, one should take $P_l=1$, then $\lambda_{1,2}=0.68$, $l_{1,2}=0.6$.
Simulation has been perfomed for a laser target thickness of 1.35 (in units of the Compton scattering length)
and the CP-IP distance $b=\gamma \sigma_y$; it gave $dL_{0,\GG}/dz/ L_{ee}= 0.84$ and 0.35 for cases 1 and 2, respectively.
The corresponding effective cross sections are 75 fb and 28.5 fb, which should be compared with 290 fb for the process $\EPEM\ \to ZH$.

The geometric $ee$ luminosity in the case of the photon collider is approximately equal to the \EPEM\ luminosity
(the pinch factor in \EPEM\ collisions is compensated by a tighter focusing in \GG\ collisions).
This means that for the same beam parameters the Higgs production rate at the photon collider
is approximately four times lower than in \EPEM\ collisions.

The photon collider can measure better only $\Gamma_{\GG}$, which determines the Higgs
production rate in \GG\ collisions and can be measured by detecting the decay mode
$H\to bb$ ($\sim57\%$ of the total number of Higgs decays).
In \EPEM\ collisions, the Higgs' \GG\
width is measured in the $H \to \GG$ decay, which has a branching fraction of
0.24\%. This means that at the photon collider the statistics for the
measurement of $\Gamma(H\to \GG)$ is higher by a factor of
$0.57/0.0024/4\approx 60$ (or even larger if a lower-emittance
electron source becomes available). This is the main motivation for the
photon collider. The study of the $H\GG$ coupling is arguably the most interesting area of Higgs
physics because it procedes via a loop and therefore is the most sensitive to New Physics.
The photon collider at the ILC with the expected $L_{ee}\approx 3\times 10^{34}$ will produce
about 22500 Higgs bosons per year ($10^7$ sec), which would enable
the determination of $\Gamma(H\to \GG)\times Br(H\to bb)$ with an accuracy of 2\%
\cite{Krawczyk,Asner,Monig}.

The photon collider can also be used also for the measurement of the Higgs boson's $CP$
properties using lineary polarized high-energy
photons (details are provided below).

As one can see, while \EPEM\ collisions are more powerful overall for the study
of Higgs properties, a \GG\ collider would add very significantly in some areas.
The relative incremental cost of adding a photon collider to an \EPEM\ linear collider
is very low. Therefore, the best solution would be to build an \EPEM\ linear collider
combined with a photon collider; the latter would come almost for free.

\subsection{The collider energy for the \GG\ Higgs factory}

 The preferable electron beam energy and laser wavelength for the \GG\ Higgs factory are
 $E_0 \approx 110$ GeV and $\lambda \approx 1\, \MKM$, corresponding to the parameter $x\approx
2$ (this includes the spectrum shift due to nonlinear effects in
Compton scattering). Note that all photon-collider projects that appeared
in the last year assumed $E_0=80$ GeV (85 GeV would be more correct)
and $\lambda=1/3$ \MKM\ ($x=4.6$). This choice was driven by the
simple desire to have the lowest possible collider energy. However, life is
not so simple, there are other important factors that must be considered:

\begin{enumerate}
\item As proposed, these projects would suffer from the very serious problem
of the removal of used electron beams. That is because the minimum energy of electrons
after multiple Compton scattering in the conversion region will be
a factor of 4.5 lower~\cite{TESLATDR}, and these electrons will be deflected
at unacceptably large angles by the opposing beam as
well as by the solenoid field (the latter due to the use of the crab-crossing collision scheme).

\item For the measurement of the Higgs' $CP$ properties one should collide
linearly polarized $\gamma$ beams at various angles between
their polarization planes. The effect is proportional to
the product of linear polarizations $l_1l_2$. The degree of linear
polarization at the maximum energies is 60\% for $x=2$ and 34.5\% at
$x=4.6$. This means that the effect in the latter case will be 3
times smaller, and so in order to get the same accuracy one would have to
run the experiment 9 times longer.
The case of $x=1.9$ was simulated, with backgrounds taken into account, in ref.~\cite{Asner};
it was found that the $CP$ parameter (a value between 1 and $-1$) can be measured with a 10\% accuracy
given an integrated geometric $ee$ luminosity of $3\cdot 10^{34} \times
10^7$ = 300 fb$^{-1}$.
\end{enumerate}
Both of these facts strongly favor a photon collider with $E_0=110$ GeV and $\lambda \approx 1$ \MKM.

\section{Photon colliders at ILC and CLIC}

The future of these collider projects is quite unclear due to their
high cost, complexity, and (as of yet) absence of new physics in
their energy region (other than the Higgs boson). If ILC in Japan
is approved, there is a very high probability that it will
include the photon collider.

The photon collider for TESLA
(on which ILC is based) was considered in detail at the conceptual
level~\cite{TESLATDR,telnov}. The next major step must be R\&D for
its laser system. Until a year ago,
the most promising solution for the laser system was an external
optical cavity, which would reduce the required laser power by a factor of
100. Such a laser system, while certainly feasible, would not be easy
to build and would require a great deal of R\&D and prototyping.
The optical-cavity technology, proposed for the photon collider in 1999,
has been developed very actively for many applications based on Compton
scattering; however, its present status is still far from what is needed
for the photon collider.

 New hopes arise from LLNL's laser-fusion project LIFE, which is based
 on the diode-pumping technology. LIFE's laser system will consist of about
 200 lasers, each operating at a repetition rate of 16~Hz and delivering
 8.4 kJ per flash. The photon collider at the ILC would require a laser
 that produces 1 ms trains of 2600 pulses, 5-10 J per pulse, with a
 repetition rate of 5-10 Hz.  LLNL experts say that the LIFE laser can
 be modified for the production of the required pulse
 trains with further chirped pulse compression. The advancement of this
 technique has been enabled by the significant reduction of the cost of
 pumping diodes, currently estimated at \$0.10 per watt,
 which translates to \$3 million per laser (the ILC-based photon collider
 would require $\sim 6$ such lasers).

 Naturally, it is very
 attractive to simply buy a few \$3M lasers and use them in one-pass mode
 rather then venturing to construct a 100 m optical cavity and stabilize
 its geometry with an accuracy of several nanometers. For the CLIC-based
 photon collider, the optical-cavity approach would not work at all due to
 CLIC's very short trains; a LIFE-type laser is therefore the only viable option.

The expected \EPEM\ luminosity of the updated ILC design at $2E_0=250$ \GEV\ is
$3\cdot 10^{34}\ \CMS$. The geometric $ee$ luminosity at the \GG\ collider
could be similar. To further increase the \GG\ luminosity, one needs
new ideas on the production of low-emittance polarized electron beams.
ILC damping rings are already close to their ultimate performance.
To increase the luminosity further, I have proposed~\cite{telnov2} to combine
many (about 50-100) low-charge, low-emittance
bunches from an RF photogun into a single bunch in the longitudinal phase
space using a small differential in beam energies.
Using this approach, it may be possible to
increase the luminosity by a factor of 10 compared to that with damping
rings. To achieve this, we need low-emittance polarized RF guns, which have
appeared only recently and are yet to reach their ultimate
performance. In the past, only DC polarized photoguns were available, which
produce beams that require further cooling with damping rings. The idea of
beam combining is highly promising and needs a more careful consideration.

The TESLA TDR, published in 2001, dedicated a 98-page chapter to
the photon collider. The recently published ILC TDR, on the other hand,
includes only a brief mention of the photon collider, as an option.
The scope document on linear colliders, developed and supported by the physics
community, states that the ILC design should be compatible with the
photon collider. The focus of the present ILC TDR was the minimization of cost
while attempting to preserve ILC's primary performance characteristics.
This has resulted in cuts in all places possible. In particular,
only one IP remains in the design, instead of two, with two pull-push detectors.
In the ILC TDR, the IP was designed for a beam crossing angle of 14 mrad,
while the photon collider requires a crossing angle of 25 mrad.
The choice of a crossing angle incompatible with the photon collider
was made simply because all attention in the TDR effort was focused on the
baseline \EPEM\ collider, not because someone was against the photon collider
(no one was). It is not too late to reoptimize the ILC IP and make it compatible
with the photon collider. Two IPs would be the best solution.

\section{Photon colliders based on recirculating linacs}
\begin{figure}[!htb]
\centering
\includegraphics[width=10.cm,height=6.5cm,angle=0]{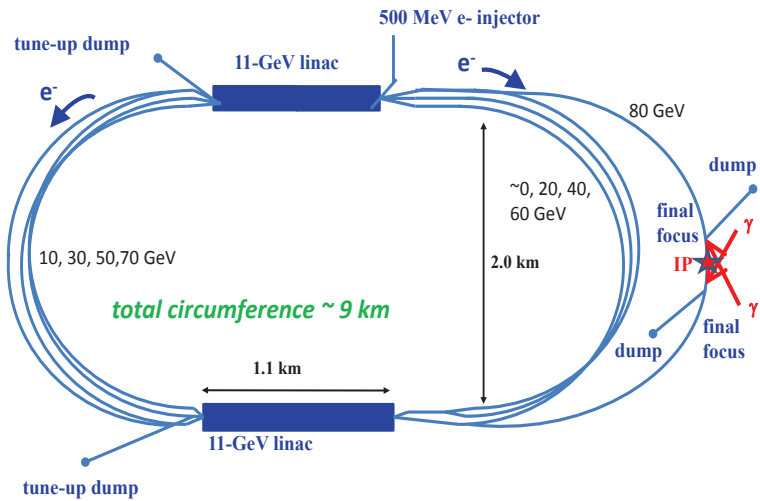}
\caption{The SAPPHiRE Higgs factory } \label{sapphire}
\end{figure}

  About one year ago, F.~Zimmermann et al.~\cite{Sapphire} proposed to use
the 60 GeV recirculating electron linac developed for $ep$ collisions with
LHC protons (LHeC) as a photon collider (project SAPPHiRE).
The ring contains two 11 GeV superconducting linacs and six arcs, each
designed for its own beam energy.
An injected electron would make three turns
to reach the energy of 60 GeV required for LHeC. To obtain the 80 GeV required
for the photon collider, the authors propose adding two additional
arcs, see Fig.~\ref{sapphire}. One must also double the number of arcs
to accomodate the
second electron beam traveling in the opposite direction. It was proposed
to use polarized electron beams with no damping rings; the required photoguns
are still under development.

  In any case, the idea is interesting because two 80 GeV electron beams
are obtained with only 22 GeV's worth of linac.
The radius of arcs is 1 km, and the total circumference is 9 km.
On the other hand, the total length of all arcs is 72 km!
In fact, about 15 years ago I considered a substantially similar
approach for a photon collider in the HERA ring at DESY
(recall that the HERA ring has four straight sections).
My conculsion was that such a design would be
impractical due to the unacceptable increase of horizontal
emittance in the bending arcs. The increase
of the normalized emittance per turn is proportional to $E^6/R^4$.
To solve this problem, the authors of SAPPHiRE have proposed to use $\times 4$
shorter arc structures, which would lead to $\times 64$ smaller emittance
dilution. This might be possible but would require $\times 16$
stronger quadrupole magnets.

  Another weak point of this proposal is the use of 80 GeV electron beams and the 1/3
  \MKM\ laser wavelength. As mentioned above,
  this choice of parameters makes it very difficult to remove the
  disrupted electron beams from the detector
 and leads to low sensitivity in the measurement of the
  $CP$ properties of the Higgs boson.

  It is highly unlikely that the LHeC project (and, correspondingly, SAPPHiRE) will be approved.
  However, the idea behind SAPPHiRE has become very popular and has been cloned for all existing
  tunnels at major HEP laboratories. In particular, it has been proposed to
  build a photon collider in the Tevatron ring at FNAL (6 km circumference),
Higgs Factory in Tevatron Tunnel (HFiTT)
~\cite{FNAL}.
  This collider would contain 8 linac sections providing a total energy gain of
  10 GeV per turn. In order to reach the energy of 80 GeV,
  the electron beams would make 8 turns. The total number of beamlines in
  the tunnel will be 16, with the total length of approximately 96 km. This
  proposal contains just a desired set of numbers without any attempt at
  justification. Simple  estimates show that such a collider will not work due to the strong
  emittance dilution both in the horizontal and vertical directions. The
  eight arcs would be stacked one on top another, so electrons will jump up and down, by up to 1.5 m, 16
  times per turn, 128 times in total. The vertical emittance is
  assumed to be same as in the ILC damping ring; it will be certainly destroyed on such ``mountains''.

\begin{figure}[!htb]
\centering
\includegraphics[width=10.cm,height=6.5cm,angle=0]{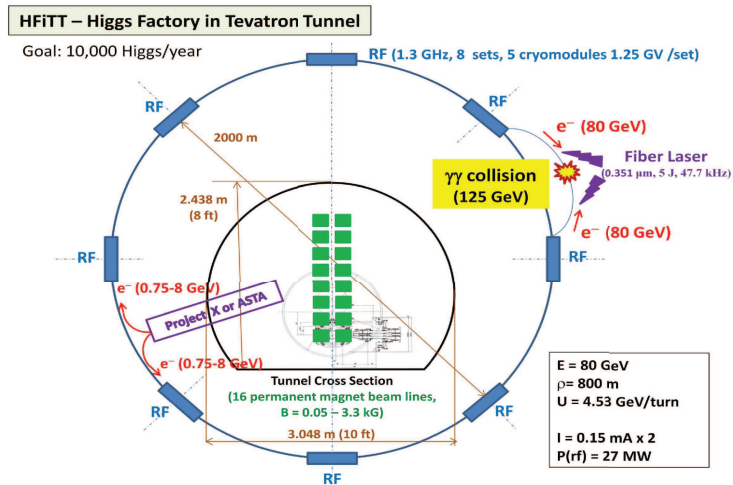}
\caption{The HFiTT Higgs factory} \label{fnal}
\end{figure}

 The most interesting feature of the HFiTT proposal is a novel laser
  system based on fiber lasers. Only recently have laser physicists
  succeeded in coherently combining the light from thousands of fibers.
  A diode-pumped fiber laser is capable of producing 5-10 J pulses with a repetition rate of 47.7 kHz
  as required by HFiTT. It would have been very attractive to use such a fiber laser for
  the photon collider at the ILC as its total power would be larger than
  needed. Unfortunately, the pulse structure at the ILC would be very bad for a such
  laser, as the ILC needs $2600\times 10$ J $=26$ kJ per 1 ms, which translates to a 55
  times greater (peak) power of the diode system. Correspondingly,
  the diode cost would be greater by the same factor.

  There is also a proposal~\cite{raubenheimer} to build a photon collider based on the existing
  SLAC linac. Electrons would acquire 40 GeV traveling in the linac in
  one direction, then make one round turn in a small ring, get another
  40  GeV traveling in the same linac in the opposing direction, and then the two
  beams would collide in $R=1$ km arcs, similar to the SLC.
  It is a nice proposal; however, for the Higgs factory it is desirable
  to have $E_0=110$ GeV, as explained above. Reaching 110 GeV would require
  either a higher acceleration gradient (or an additional
  30 GeV injector) and arcs with a larger radius.

\section{Conclusion}

  The photon collider based on ILC (or CLIC) is a highly realistic
  project. However, if the \EPEM\ program occupies all the experiment's time,
  the photon collider will not become reality for least 40 years from now, which is
  unattractive for the present generation of physicists. The best
  solution for this problem is to build a collider with two interaction regions.

  A laser system based on the project LIFE lasers is the most attractive choice at this time;
  fiber lasers can also reach the desired parameters at some point in future.
  Development of low-emittance polarized electron beams can
  increase the photon collider luminosity by a further order of
  magnitude. The photon collider would be very useful for the precise measurement
  of the Higgs' \GG\ partial width and its $CP$ properties. A very high-luminosity photon
  collider at the energy $2E_0=400$ GeV can help measure the Higgs' self coupling. The
  photon collider based on ILC (CLIC) can work with the 1 \MKM\ laser
  wavelength up to $2E_0 \sim 700$ GeV; for higher energies, one should use a greater laser wavelength.

  The idea of a photon-collider Higgs factory based on recirculating linacs
  looks interesting as it can use shorter linacs. Unfortunately, the problem of
  emittance dilution is very serious and the total length of the arcs is very large.
  The pulse structure of such colliders (equal distance between collisions) is very well suited for fiber
  lasers. Such a recirculating collider with a desirable $E_0$ $(\approx 110$ GeV) can possibly work in large rings such as
  LEP/LHC or UNK, but then the total length of arcs will be several hundred km and the cost would exceed
  that for linear colliders with similar energy (that could be, for example, a warm linear collider with the 4 km length).
  Most importantly, a photon collider with no \EPEM\ does not make much sense for the study of the Higgs boson.
  At this time, the ILC is the best place for the photon collider.

\end{document}